\documentclass[aps,twocolumn,superscriptaddress,floatfix,letter]{revtex4}

\usepackage{graphicx}
\usepackage{amsmath}
\usepackage{hyperref}
\usepackage{url}
\usepackage{float}

\usepackage{color}

\newcommand{\ket}[1]{\left\vert #1\,\right\rangle}
\newcommand{\bra}[1]{\left\langle #1\,\right\vert}

\begin{document}

\setlength\abovedisplayskip{5pt}
\setlength\belowdisplayskip{5pt}

%\title{Emergence of the Bose-Einstein distribution and correlations between occupation %numbers in quench dynamics}

\title{Emergence of correlations in the process of thermalization of interacting bosons}

\author{Fausto Borgonovi}
\affiliation{Dipartimento di Matematica e
  Fisica and Interdisciplinary Laboratories for Advanced Materials Physics,
  Universit\`a Cattolica, via Musei 41, 25121 Brescia, Italy}
\affiliation{Istituto Nazionale di Fisica Nucleare,  Sezione di Pavia,
  via Bassi 6, I-27100,  Pavia, Italy}
\author{Felix M. Izrailev}
\affiliation{Instituto de F\'{i}sica, Benem\'{e}rita Universidad Aut\'{o}noma
  de Puebla, Apartado Postal J-48, Puebla 72570, Mexico}
\affiliation{Dept. of Physics and Astronomy, Michigan State University, E. Lansing, Michigan 48824-1321, USA}

\date{\today}

\begin{abstract} We address the question of the relevance of thermalization to the increase of correlations in the quench dynamics of an isolated system with a finite number of interacting bosons. Specifically, we study how, in the process of thermalization, the correlations between occupation numbers increase in time resulting  in the emergence of the Bose-Einstein distribution. We show, both analytically and numerically, that before saturation the two-point correlation function increases quadratically in time. This time dependence is at variance with the exponential increase of the number of principal components of the wave function, recently discovered and explained in Ref.\cite{BIS18}. We also demonstrate that the out-of-time-order correlator (OTOC) increases algebraically  in time but not exponentially as predicted in many publications. Our results, that can be confirmed experimentally in traps with interacting bosons,  may be also relevant to the problem of black hole scrambling.
\end{abstract}

\pacs{05.30.-d, 05.45.Mt, 67.85.-d}
\maketitle

{\it Introduction -} In recent years the problem of thermalization in closed systems of interacting fermions and bosons has attracted much attention (see, for example, Refs.\cite{reviews,BISZ16}). An increase of interest to this problem is due to remarkable experimental achievements \cite{exp} and various theoretical predictions \cite{zele,flam,deutsch}. Although the term {\it thermalization} is not uniquely defined, it is widely used in many-body physics. One of the basic statistical properties of many-body systems is either the Bose-Einstein (BE) or Fermi-Dirac (FD) distribution that emerge in the thermodynamic limit due to the combinatorics and  without inter-particle interaction. As for finite isolated systems, the mechanism for the onset of  BE and FD distributions is the chaotic structure of many-body eigenstates \cite{zele,flam,BGIC98,BCIC02,BISZ16,BMI17}. In this case, the interaction between particles plays a crucial role: the fewer the particles the stronger the inter-particle interaction has to be for the emergence of the statistical properties. 

To date it is understood that the validity  of statistical mechanics can be justified not only by averaging over a number of eigenstates with close energies, but also with the use of a single eigenstate if the latter consists of many uncorrelated components in the physically chosen basis. Specifically, it was shown that BE and FD distributions emerge also on the level of individual eigenstates if they are strongly chaotic \cite{BGIC98,GGF99,BMI17}. The most intriguing point here is that both distributions appear even if the number of particles is small; this happens due to the fast growth of the number of components in many-body eigenstates in dependence on the number of particles.  

Unlike the onset of BE and FD distributions emerging from single stationary eigenstates, in this Letter we address a new problem concerning the onset of the BE distribution in the evolution of a system with few interacting bosons. Our specific interest is to study how the conventional BE distribution emerges {\it in time} and how this fact is related to the somewhat different  problem of the increase of correlations in the process of relaxation of a system to a steady-state distribution. The latter problem is now a hot topic in  literature in view of various applications, such as the evolution of systems with cold atoms, as well as in application to the problem of scrambling in black holes (see \cite{M18} and references therein). 

In our study we consider the quench dynamics described by the Hamiltonian $H=H_0+V$ where $H_0$ represents the non-interacting bosons and the interaction is fully embedded into $V$ belonging to the ensemble of two-body random interacting (TBRI) matrices. In this way, by exciting initially a single many-body state of $H_0$ we explore the evolution of wave packets in the Fock space. Recently, it was discovered that for the model parameters for which the many-body eigenstates of $H$ are strongly chaotic, the effective number of components $N_{pc}$ in the wave function increases exponentially in time, before the saturation which is due to the finite number of particles \cite{BIS18}. This time dependence was explained with the use of a phenomenological model that  allowed to obtain simple analytical expressions for the rate of exponential increase of $N_{pc}$ and for its saturation value.   

Below, in connection with the results reported in \cite{BMI17,BIS18} we show, both analytically and numerically, that the onset of the BE distribution in the TBRI matrix model occurs on the time scale on which the number of components in many-body eigenstates increases exponentially in time. In order to quantify the onset of the BE distribution we have studied the correlations between occupation numbers by exploring both the two- and four- point correlators. The latter is just the well known OTOC correlator widely discussed in literature \cite{OTOC}. Specifically, it was predicted that for strongly chaotic systems  OTOC should manifest an exponential time-dependence before saturation. One of our main findings is that actually both correlators increase algebraically in time  and not exponentially. This result is quite unexpected,  as compared with the exponential increase of the number  of principal components $N_{pc}$ in the wave packet. Our analytical results are fully confirmed by extensive numerical data.
     
{\it The model -} The system consists of $N$ identical bosons occupying $M$ single-particle levels specified by random energies $\epsilon_s$ with mean spacing, $\langle \epsilon_s- \epsilon_{s-1} \rangle = 1 $. The Hamiltonian $H= H_0 + V $ reads ($\hbar=1$),
\begin{equation}
  H= \sum \epsilon_s \, a^\dag_s a_s  +
 \sum V_{s_1 s_2 s_3 s_4} \, a^\dag_{s_1} a^\dag_{s_2} a_{s_3} a_{s_4}
\label{ham}
\end{equation}
where the two-body matrix elements $ V_{s_1 s_2 s_3 s_4} $ are random Gaussian entries with  zero mean and variance $V^2$. The dimension of the Hilbert space generated by the many-particle basis states is ${N_H} = (N+M-1)!/N!(M-1)!$ Here we consider  $N=6$ particles in $M=11$ levels (dilute limit, $N \leq M$) for which $N_H = 8008$. Two-body random  matrices (\ref{ham}) were introduced  in \cite{TBRI,brody}  and extensively studied for fermions \cite{flam,alt} and bosons \cite{bosons}. 

The eigenstates $\ket{\alpha} = \sum_k C_k^{(\alpha)} \ket{k}$ of $H$ can be written in terms of the basis states $|k\rangle = a^\dagger_{k_1}...a^\dagger_{k_N} |0\rangle$ of $H_0$, where
\begin{equation}
H |\alpha \rangle = E^\alpha |\alpha \rangle ; \quad H_0 |k\rangle = E^0_{k} |k\rangle .
\label{ldos}
\end{equation}
An eigenstate $\ket{\alpha}$ of the total Hamiltonian is called chaotic when its number $N_{pc}$ of  principal components 
$C_k^{\alpha}$  is sufficiently large and $C_k^{\alpha}$ can be considered as random and non-correlated ones. Note that since the system is isolated and the perturbation $V$ is finite, the eigenstates can fill only a part of the unperturbed basis~\cite{BISZ16} determined by the perturbation $V$. Specifically, the energy region which is occupied by the eigenstates is restricted by the width of the so-called energy shell~\cite{chirikov}. The partial filling of the energy shell by an eigenstate can be associated with the many-body localization in the energy representation. Contrary, when an eigenstate fills completely the energy shell, we are in presence of  maximal quantum chaos, and the BE distribution  emerges on the level of individual eigenstates~\cite{BGIC98,GGF99,BMI17}. This happens when the interaction $V$ is sufficiently large, $V > V_{cr}$, to provide strong quantum chaos. In what follows we will consider the situation when the latter condition is fulfilled. 

{\it Dynamics in Fock space -} In contrast with the previous studies~\cite{BMI17}, focused on the  thermal properties of individual many-body eigenstates, here we consider the dynamics of the model (\ref{ham}) by exploring two different time scales, before and after the relaxation to a steady state. Specifically, we study the quench dynamics starting from a single many-body state $\ket{k_0}$ of the unperturbed Hamiltonian $H_0$, after  switching on the interaction $V$. Given the evolved wave function
$\ket{\psi(t)} =  e^{-iHt} \ket{k_0}$  one can express the probability $P_k(t) =  |\bra{k} \psi(t) \rangle|^2$ to find the system at time $t$ in any unperturbed state $\ket{k}$ as follows,
\begin{equation}
P_k(t) = \sum_{\alpha,\beta}  C_{k_0}^{\alpha \ast} C_{k}^{\alpha } C_{k_0}^\beta C_{k}^{\beta \ast} e^{-i(E^\beta-E^\alpha )t} 
\equiv P_{k,k_0}^d + P_{k,k_0}^f (t),
\end{equation}
where $P_{k,k_0}^d=\sum_{\alpha}  |C_{k_0}^\alpha|^2 |C_{k}^\alpha|^2$ and $P_{k,k_0}^f (t)$ are the time-independent and time-fluctuating parts, respectively.
With this expression, one can analyze the number of principal components,
\begin{equation}
\label{s-ipr}
 N_{pc}(t) = \left\{ \sum \limits_{k} \left[P_{k,k_0}^d + P_{k,k_0}^f (t)\right]^2  \right\}^{-1},
\end{equation}
known as the participation ratio.  
Taking the long-time average, $P_{k,k_0}^f(t)$ cancels out  and only the diagonal part $P_{k}^d$ survives. As is shown in \cite{BIS18}  the number $N_{pc}(t)$ of principal components in the wave packets increases {\it exponentially fast} in time, $N_{pc}(t) \sim\exp(2\Gamma t)$ up to some saturation time $t_s$. The rate of the exponential growth  is defined by the width $\Gamma$ of the local density of states (LDOS), 
$$F_{k_0} (E) =\sum_{\alpha}  | C_{k_0}^{\alpha}|^2 \delta (E - E^{\alpha }), $$
 obtained by projecting the initial state $\ket{k_0}$ onto the energy eigenstates. In nuclear physics it is known as 
{\it strength function}  and   it  describes the relaxation of  excited heavy nuclei~\cite{bohr}.
Concerning  the saturation time $t_s$,  it was found~\cite{BIS18} to be proportional to the number of particles, $t_s \approx N/\Gamma$. This time should be treated as the time after which one can speak of a {\it complete thermalization} occurring in a system.  The exponential increase of $N_{pc}(t)$ is shown in \cite{SM}, together with the analytical estimates obtained in Ref.~\cite{BIS18}.

%%%%%%%%%%%%%%%%%%%%%%%%%%%%%%%%%%%%%%%%%%%%%%%%%%%%%%%%%%%%%%%%%%%%%%%%%%%%%%%%
\begin{figure}[t]
\vspace{0.cm}
\includegraphics[width=\columnwidth]{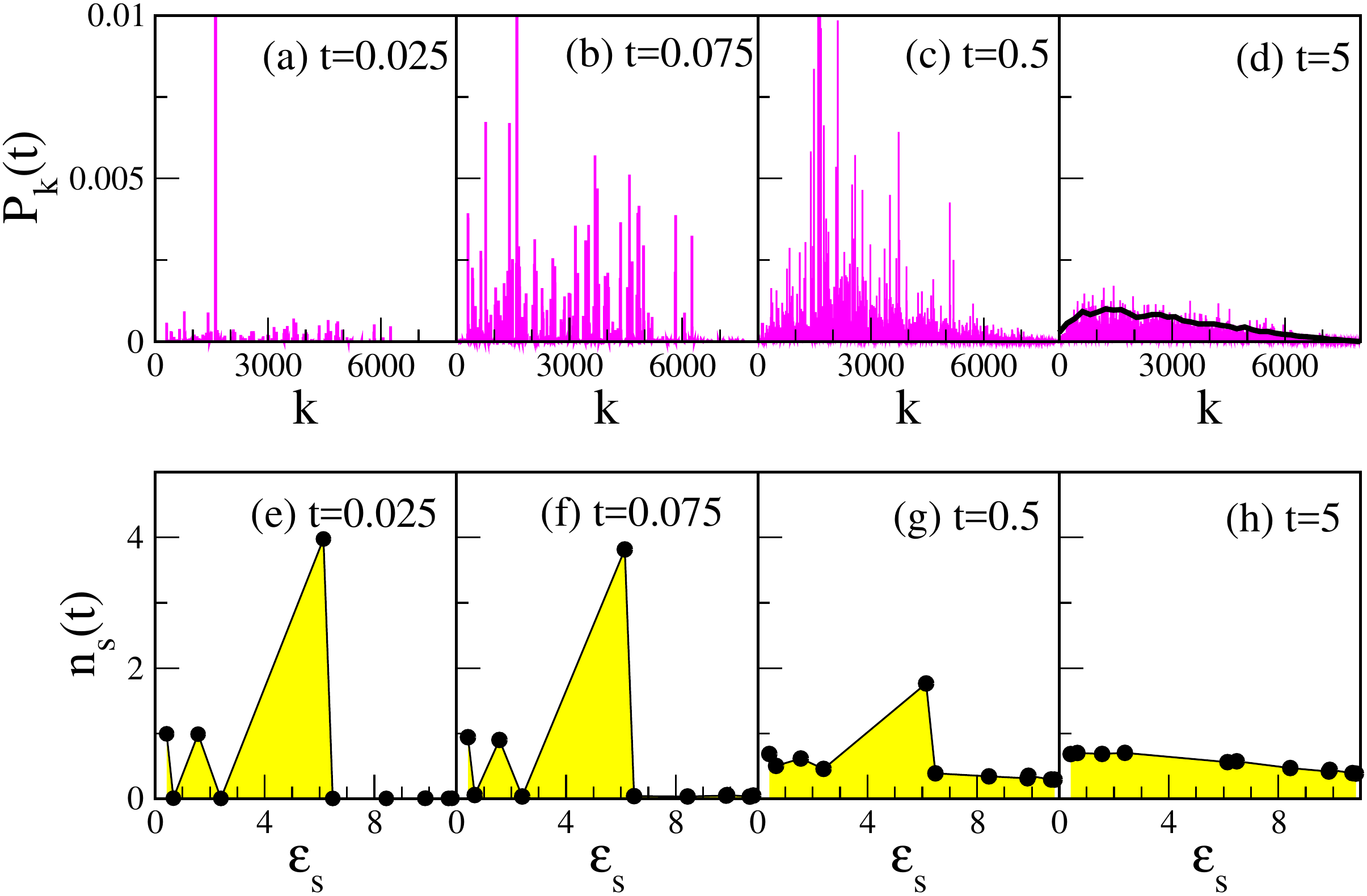}
\caption{(Color online) Upper panels: Probability $P_k(t)$ at different times $t$ in the unperturbed basis $|k\rangle$. Low panels show $n_s(t)$ versus the single-particle energies $\epsilon_s$. In panel (d) the envelope of the stationary distribution is shown by a black curve. Initial state is $\Psi_0 = |1 0 1 0 4 0 0 0 0 0 0\rangle$ where integer numbers are numbers of bosons occupying the $s$-level. Dynamics is shown for $N=6, M=11$ and $V=0.4$. For this value of $V$ the eigenstates are strongly chaotic~\cite{BMI17}.
}
\label{f-01}
\end{figure}
%%%%%%%%%%%%%%%%%%%%%%%%%%%%%%%%%%%%%%%%%%%%%%%%%%%%%%%%%%%%%%%%%%%%%%%%%%%%%%%%

{\it Onset of Bose-Einstein distribution -} The time-dependent occupation number distribution (OND) is defined as follows, 
\begin{equation}
\label{ond-1}
n_{s} (t)  = \bra{\psi(t)} \hat{n}_s  \ket{\psi(t)} =  \sum_ k n_s^k  |\langle k |\psi (t) \rangle |^2 .
\end{equation}
It gives the average number of particles in the single-particle energy level $\epsilon_s$ at the time $t$. Here we took into account that 
$\bra{k} \hat{n}_s \ket{k^\prime} = n_s^k \, \delta_{k,k^\prime}  $ where $n_s^k =0,...,N$. The evolution of $n_s(t)$ in comparison with the wave packet dynamics $P_k(t)$ is shown in Fig~\ref{f-01}(e)-(h). This figure demonstrates that when the packet fully occupies the energy shell, the occupation numbers are relaxed to the steady-state distribution. 

Expanding $e^{-iHt}$ at second order one gets the time dependence for $n_s(t)$ at small times, 
\begin{equation}
 |\langle k | e^{-iHt} | k_0\rangle |^2 \simeq \delta_{k,k_0} +t^2 \left[ H_{k,k_0}^2 - \delta_{k_0,k_0} (H^2)_{k,k_0}\right] 
+o(t^4)
\label{noc-as1}
\end{equation}
which results in the following estimate,
\begin{equation}
n_{s}(t)  \simeq  n_s^{k_0} +  t^2  \sum_ {k\ne k_0} (n_s^{k_0}-n_s^k)   H_{k,k_0}^2 + o(t^4)
\label{noc-st}
\end{equation}
One can see in Fig.~\ref{f-03} that for single-particle $s-$levels which are not initially occupied by particles,   $n_s(t)$ grows quadratically in time. As for the saturation values $\overline{n_{s}}$ after the relaxation time $t_s$, they can be also obtained analytically by performing an infinite time average,
\begin{equation}
\overline{n_{s}} =  \sum_k n_s^k  \overline{|\langle k |\psi (t) \rangle |^2 } = 
\sum_k n_s^k P_{k,k_0}^d \ .
\label{noc-as}
\end{equation}
In order to claim that after relaxation the OND is statistically  described by a  BE distribution,
 one has to be sure that the fluctuations of $n_s$ follow the standard requirements of statistical mechanics. 
In view of this very point, we have thoroughly analyzed both ``classical'' and ``quantum'' fluctuations. Concerning the former, they can be analyzed by the search of the time dependence of $n_s(t)$ with respect to their asymptotic values reached after relaxation. According to the statistical mechanics, a) the fluctuations have to be small as compared to the mean values, and b)   fluctuations should be   Gaussian.    Our numerical analysis of the fluctuations, see \cite{SM},  has shown  that the relative fluctuations $\Delta n_s / \langle n_s \rangle $ are Gaussian distributed  and   decreasing  as $1/\sqrt{\overline{N}_{pc}}$ (infinite time average of $N_{pc}(t)$) instead of  $1/\sqrt{N}$ (number of particles).  This remarkable result  shows that for systems having few chaotic interacting particles the number of principal components 
$\overline{N}_{pc}$
in the wave packet plays the same role as the number of particles $N$ in ordinary statistical mechanics.
A more intriguing point concerns quantum fluctuations. It is a textbook result~\cite{huang} that BE statistics is characterized 
by relative quantum fluctuations 
$\delta n_s^2/n_s^2 = 1+ 1/n_s $, where   
$\delta n_s^2 = \overline{n^2}_s - \overline{n}_s^2 $ with the overbar standing for the infinite time-average, see Eq.~(\ref{noc-as}).
Once again we checked that, provided the time-dependent wave function is chaotic, fluctuations follow the predictions of standard statistical mechanics (for details see \cite{SM}).  
 This should be considered as an additional proof of the statistical character of the evolution of the system after the relaxation.

%%%%%%%%%%%%%%%%%%%%%%%%%%%%%%%%%%%%%%%%%%%%%%%%%%%%%%%%%%%%%%%%%%%%%%%%%%%%%%%%
\begin{figure}[t]
\vspace{0.cm}
\includegraphics[scale=0.45]{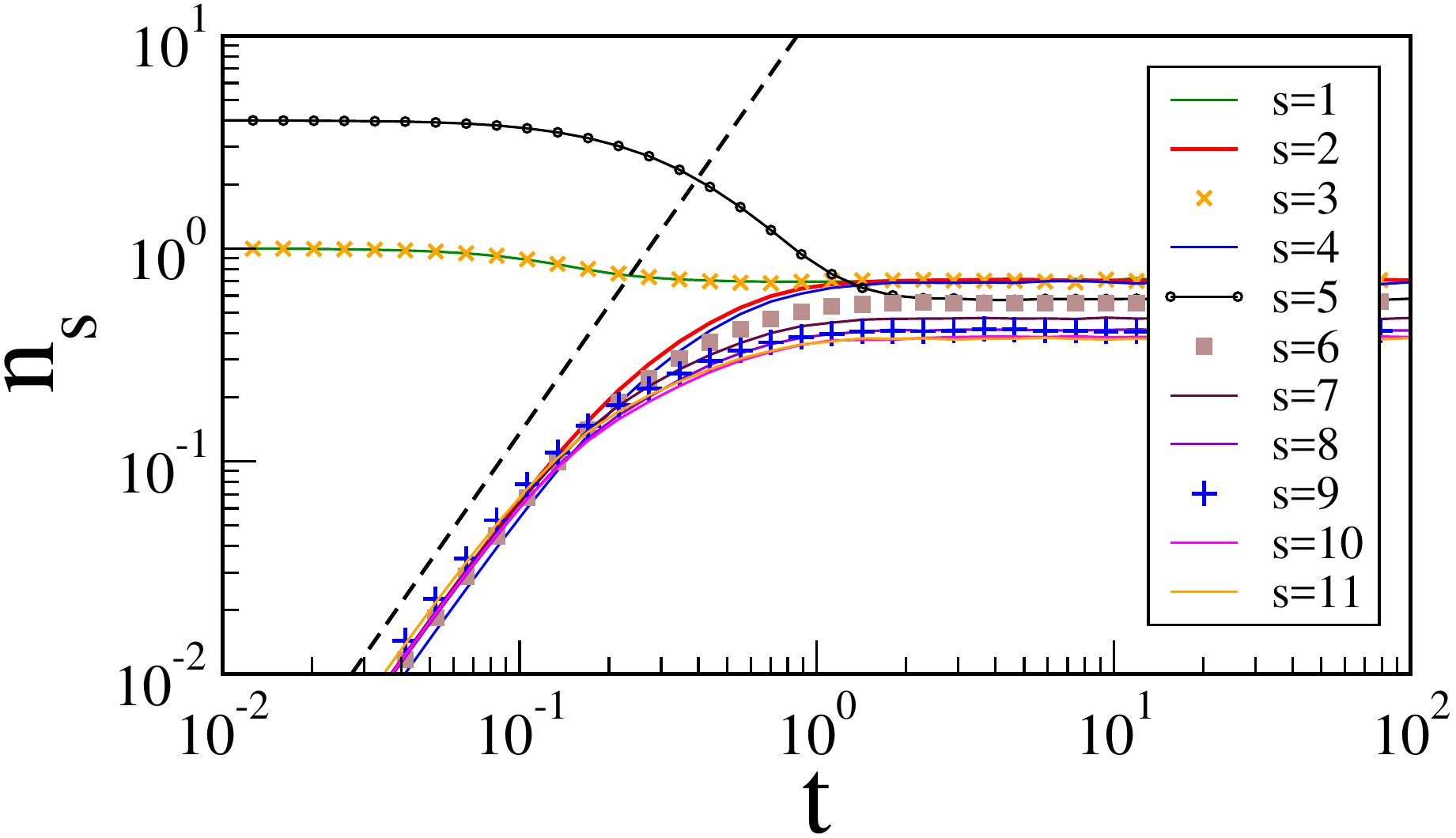}
\caption{Evolution of the averaged $n_s (t)$ for all $s=1,...,M$. Dashed line is the predicted $t^2$ behavior (\ref{noc-st}) characteristic of the perturbative regime. Initial state is $\Psi_0 = |1 0 1 0 4 0 0 0 0 0 0\rangle$. Here $N=6, M=11$ and $V=0.4$ as in Fig.\ref{f-01}. An average over 10 realizations of the random potential has been used.
}
\label{f-03}
\end{figure} 
%%%%%%%%%%%%%%%%%%%%%%%%%%%%%%%%%%%%%%%%%%%%%%%%%%%%%%%%%%%%%%%%%%%%%%%%%%%%%%%%

{\it Two-point correlation function -} Let us now study how the onset of the BE distribution is manifested by the emergence of  correlations between occupation numbers. First, we start with the two-point correlation function ${C}_{s,s+1} (t)$ between neighboring occupation numbers,
 \begin{equation}
{C}_{s,s+1} (t) = \langle k_0 | [ \hat{n}_s(t) - \hat{n}_{s} ][ \hat{n}_{s+1}(t) - \hat{n}_{s+1}] | k_0 \rangle .
 \label{corr11}
\end{equation}
Initially the correlations are absent, ${C}_{s,s+1} (0)= 0$, however, they appear in time. The time-dependence of ${C}_{s,s+1} (t) $ is shown in  Fig.~\ref{f-04} for all $s=1,..,M-1$.  As one can see, there is a clear relaxation to steady-state values after the critical time $t_s$. The negative or positive sign of the asymptotic correlations is related to the particular choice of the initial state. 

It is also instructive to introduce the global correlator $ {\cal C}^{(2)}$ which is the sum of the correlators between all neighboring single-particle energy levels $\epsilon_s$ and $\epsilon_{s+1}$, 
 \begin{equation}
{\cal C}^{(2)}(t) = |\sum_{s=1}^{M-1}   C_{s,s+1} (t)| .
 \label{corr11t}
\end{equation}
This correlator is independent of the specific $s$ level  and it can be used as a global measure of correlations between occupation numbers of nearest single-particle energy levels. Performing an expansion on a small time scale it is possible to show that
\begin{equation}
{\cal C}^{(2) } (t) \simeq
 t^2     | \sum_{s=1}^{M-1} \sum_{r=s+1}^M   \sum_k  H_{k,k_0}^2 W_{k,k_0}^{sr} | + o(t^4) 
\label{cor-t2}
\end{equation}
with $W_{k,k_0}^{sr} = [n_s^k n_{r}^k +  n_s^{k_0} n_{r}^{k_0} -n_s^{k_0} n_{r}^{k}-n_s^{k} n_{r}^{k_0} ] $. 
As one can see,  Eq.~(\ref{cor-t2}) does not contain eigenvalues and eigenfunctions. This means that 
  in order to get the initial spread of the correlator, there is no need to  diagonalize the Hamiltonian. Concerning the saturation value, it can be obtained by performing the time average for $t \geq t_s$ (see \cite{SM}),
 \begin{equation}
\overline{{\cal C}^{(2) }}   =  | \sum_{s=1}^{M-1} \sum_{r=s+1}^M   \sum_k  P_{k,k_0}^d  W_{k,k_0}^{sr}|.
\label{cor-ave}
\end{equation} 
The time evolution for ${\cal C}^{(2) } (t) $ is shown in Fig.\ref{f-04}, together with the analytical predictions. The correspondence between numerical data and analytical predictions is impressive. Thus, the dynamics of $ {\cal C}^{(2) } (t)$ is fully described by the  analytical expressions (\ref{cor-t2}) and (\ref{cor-ave}).

%%%%%%%%%%%%%%%%%%%%%%%%%%%%%%%%%%%%%%%%%%%%%%%%%%%%%%%%%%%%%%%%%%%%%%%%%%%%%%%%
\begin{figure}[t]
\vspace{0.cm}
\includegraphics[scale=0.43]{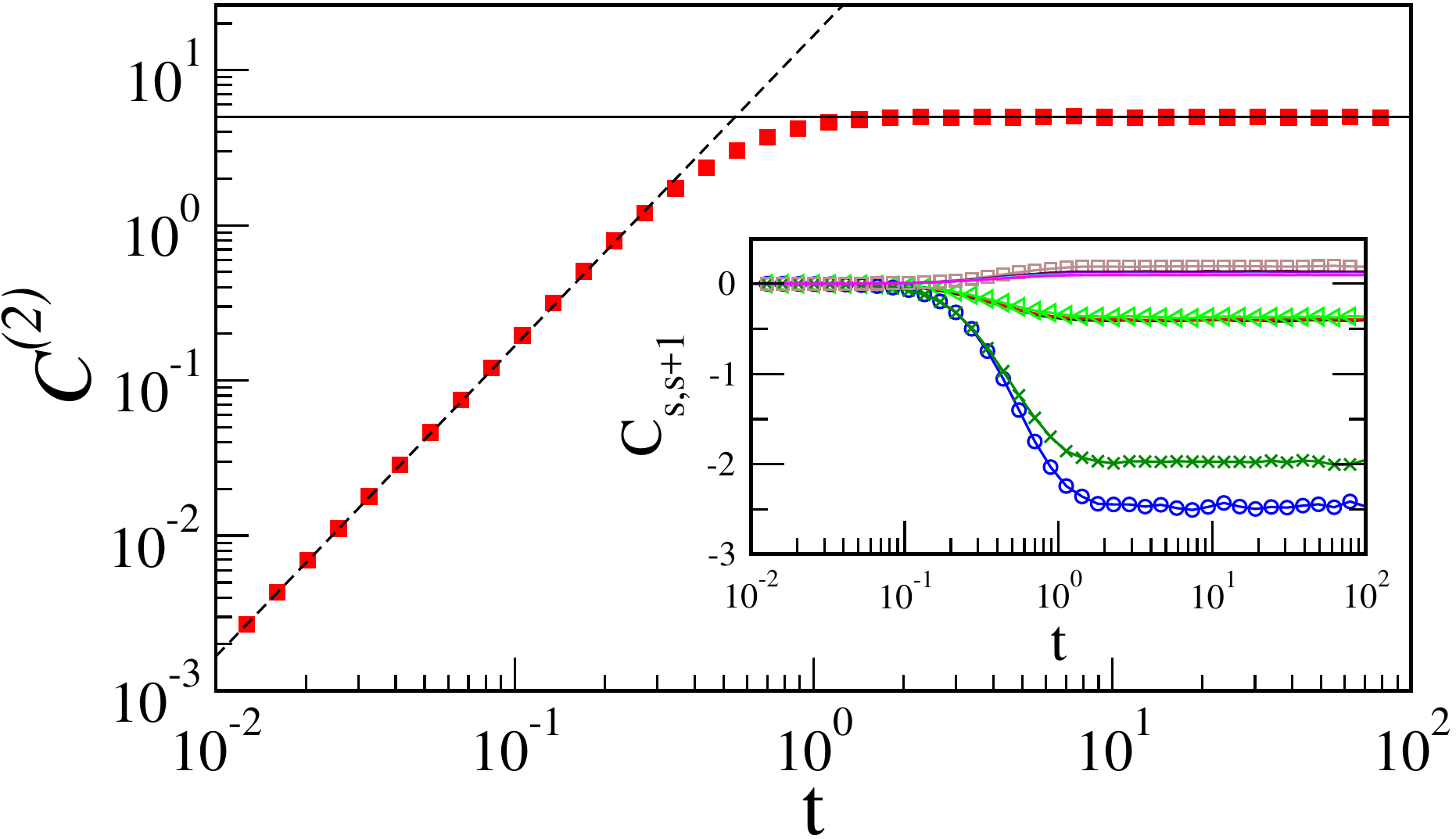}
\caption{ Global two-point correlation function ${\cal C}^{(2) } (t) $ (red squares). Dashed line is given by Eq.~(\ref{cor-t2}). Horizontal line corresponds to Eq.~(\ref{cor-ave}). The initial state and parameters are the same as in Fig.\ref{f-03}. The average over $10$ realizations of the random  potential was used. Inset: Correlation function $C_{s,s+1} (t)$ for all $s=1,...,M-1$.
}
\label{f-04}
\end{figure}
%%%%%%%%%%%%%%%%%%%%%%%%%%%%%%%%%%%%%%%%%%%%%%%%%%%%%%%%%%%%%%%%%%%%%%%%%%%%%%%%

{\it Four-point correlation function (OTOC) -} Now let us study the four-point correlator between nearest single-particle energy levels,
\begin{equation}
{\cal O}_{s,s+1} (t) = \langle k_0 |[ \hat{n}_s(t), \hat{n}_{s+1}(0)] |^2 | k_0 \rangle .
\label{otoc1}
\end{equation}
This correlator, also known  as OTOC, has been recently introduced in the frame of the SYK model \cite{SYK} and widely discussed in view of various physical applications (see e.g. \cite{OTOC}). 

After some algebra~\cite{SM}, one can obtain that the correlator ${\cal O}_{s,s+1} (t)$ increases in time quadratically on a small time scale, whose validity defines the perturbative regime,
\begin{equation}
{\cal O}_{s,s+1} (t) \simeq t^2  \sum_{k\ne k_0 }  H_{k,k_0}^2  \left(
n_s^k -  n_s^{k_0}   \right)^2 \left(n_{s+1}^k -  n_{s+1}^{k_0}   \right)^2 .
\label{otoc-t2}
\end{equation}
In the same way, by performing an infinite time-average,  we can obtain the steady state 
 value $\overline{{\cal O}_{s,s+1}  }$,
\begin{equation}
\begin{array}{lll}
&\overline{{\cal O}_{s,s+1}  } =    \sum_k \left(n_{s+1}^k -n_{s+1}^{k_0}\right)^2 
\left\{  \left[\sum_{\alpha} C_k^\alpha  C_{k_0}^\alpha {\cal N}_s^{\alpha,\alpha}\right]^2+ \right.\\
&\left. \sum_{\alpha \ne \beta} |C_k^\alpha|^2  |C_{k_0}^\beta|^2 \left( {\cal N}_s^{\alpha,\beta} \right)^2 \right\}
\label{otoc-ave}
\end{array}
\end{equation}
with ${\cal N}_s^{\alpha,\beta} = \sum_{k} C_k^\alpha  C_{k}^\beta n_s^k$.
%%%%%%%%%%%%%%%%%%%%%%%%%%%%%%%%%%%%%%%%%%%%%%%%%%%%%%%%%%%%%%%%%%%%%%%%%%%
\begin{figure}[t]
\vspace{0.cm}
\includegraphics[width=\columnwidth]{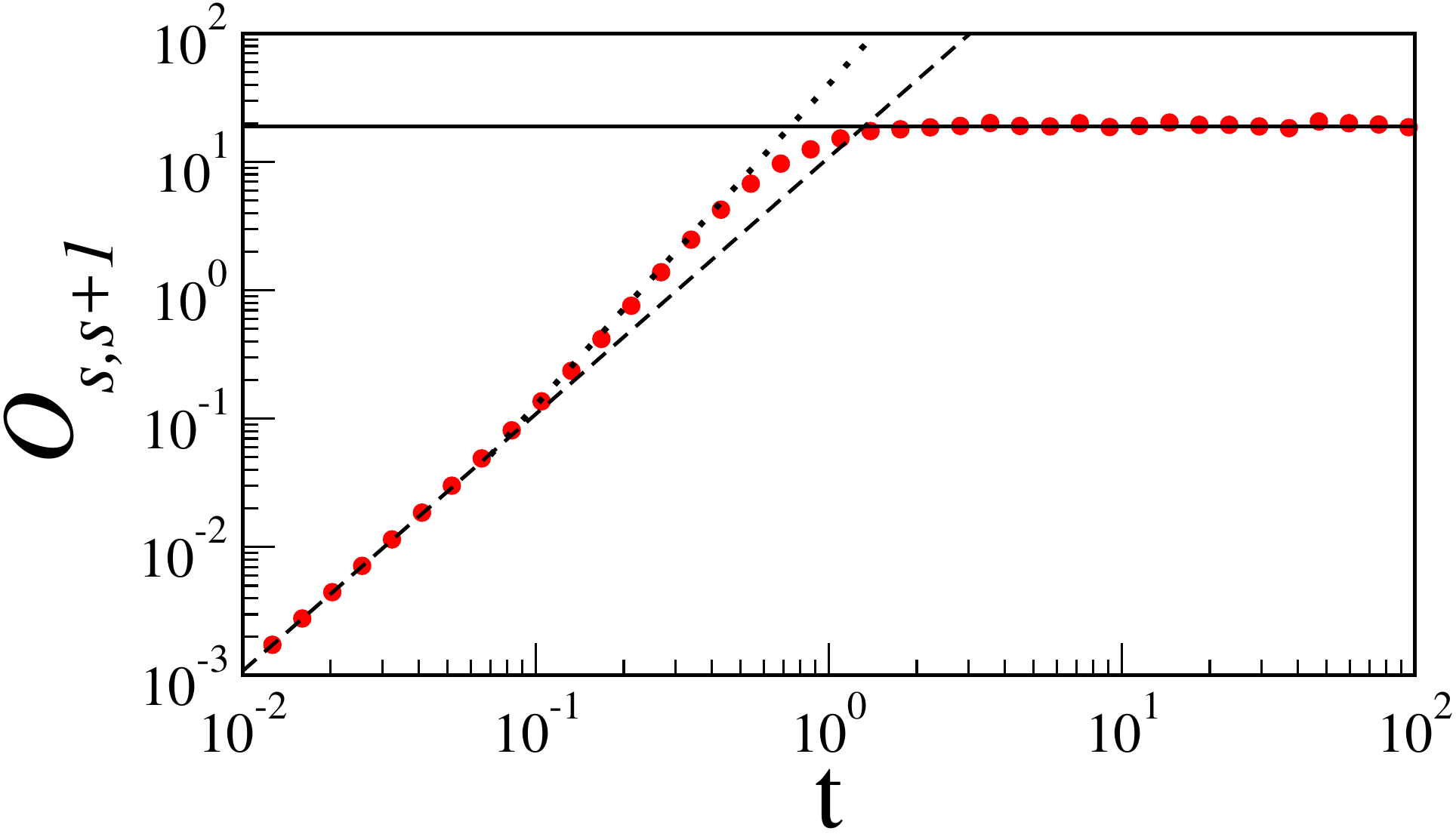}
\caption{Evolution of the four-point correlator $ {\cal O}_{s,s+1} (t)$ for $s=5$.
Dashed line is the analytical prediction (\ref{otoc-t2}). Horizontal line corresponds to Eq.~(\ref{otoc-ave}). Dotted line is the fit for $t>0.07$ (outside perturbative regime), giving the  $t^{2.5}$ dependence. Initial state is $\Psi_0 = |0 0 0 0 6 0 0 0 0 0 0\rangle$ and $N=6, M=11, V=0.4$.
}
\label{f-05}
\end{figure}
%%%%%%%%%%%%%%%%%%%%%%%%%%%%%%%%%%%%%%%%%%%%%%%%%%%%%%%%%%%%%%%%%%%%%%%%%%%%

Numerical data for ${\cal O}_{s,s+1} (t)$ are shown in Fig.~\ref{f-05} together with the expressions (\ref{otoc-t2}) and (\ref{otoc-ave}). Our results demonstrate that while in the perturbative regime the growth is indeed quadratic, a time window can be found
where the correlator  increases   approximately as  $t^{2.5}$, before the saturation. This occurs at variance with the behavior of the two-point correlator for which only the quadratic regime before saturation is seen  
and with $N_{pc}$ which grows exponentially in time.

{\it Conclusion and discussion -} In this Letter we address the question of how the conventional Bose-Einstein distribution emerges in an isolated system with a finite number of interacting bosons. Since this process is accompanied by an increase of strong correlations between occupation numbers $n_s(t)$, the large part of our study is devoted to the details of the time dependence of these correlations.  

For our analysis we have used the well known model (\ref{ham}) describing bosons interacting to each other via two-body random matrix elements. By exploring the quench dynamics, we show that   the BE distribution emerges  on the same time scale $t_s$ 
on which the number of principal components in the wave function increases exponentially in time in the Fock space\cite{BIS18}.
 This time scale $t_s$ is proportional to the number $N$ of bosons and defines the time after which one can speak of a complete  thermalization in the system. 

In order to confirm the true statistical behavior of the occupation numbers, we have carefully studied the fluctuations of $n_s(t)$ after the relaxation. In accordance with the standard statistical mechanics our data manifest that the fluctuations are of the Gaussian type, and that they are small compared to the mean values of $n_s(t)$. It was also shown that relative quantum fluctuations, $\delta n_s^2/n_s^2 $, are also in agreement with the Bose statistics (see \cite{huang}).    

In order to reveal how the process of thermalization is related to the onset of correlations, we have studied, both analytically and numerically, two correlators. One is the standard two-point correlator between nearest occupation numbers $n_s$ and $n_{s+1}$ and the other is the out-of-time order correlator (OTOC) recently discussed in literature. We have found that the two-point correlator increases in time quadratically before the saturation. As for the OTOC, initially, it also increases quadratically, however, before saturation our numerical data demonstrate the dependence $\sim t^{2.5}$ at variance with the quadratic increase predicted analytically. This result 
contradicts  the prediction  that the OTOC typically increases exponentially on some time scale \cite{OTOC}.

Our results show how the information initially encoded in a local unperturbed state, spreads over the whole system and transforms onto global correlations specified by the BE distribution of occupation numbers. Although the dynamics is completely reversible due to the unitarity of the evolution operator,  it is  {\it practically}   impossible to extract the information about the initial state, by measuring the correlations between the components of the wave function.  
Indeed the  full information about the  initial state   can be extracted only if there is an  additional  complete knowledge of the random operator     $V$ .  
Thus one can indeed speak of  the loss of information due to scrambling.  The process of this loss is accompanied by the emergence of global (thermodynamical) correlations, as  demonstrated by the data reported in this Letter.

We hope that our study can help to understand the relation between thermalization and scrambling from one side, and the onset of correlations in the evolution of chaotic systems from the other one. Since the TBRI matrix model (\ref{ham}) has been proved to manifest generic statistical properties occurring in realistic physical systems (see, for example, \cite{lieb}), the obtained results can be confirmed experimentally by studying interacting bosons in optical traps. Our results may be also important in view of the problem of black hole scrambling, see \cite{M18} and references therein.  

{\em Acknowledgements.}--
We acknowledge financial support from  VIEP-BUAP Grant No. IZF-EXC16-G (FMI) and Iniziativa Specifica INFN-DynSysMath (FB).

 \title{Supplemental Material for : Emergence of correlations in the process of thermalization of interacting bosons}

\author{Fausto Borgonovi}
\affiliation{Dipartimento di Matematica e
  Fisica and Interdisciplinary Laboratories for Advanced Materials Physics,
  Universit\`a Cattolica, via Musei 41, 25121 Brescia, Italy}
\affiliation{Istituto Nazionale di Fisica Nucleare,  Sezione di Pavia,
  via Bassi 6, I-27100,  Pavia, Italy}
\author{Felix M. Izrailev}
\affiliation{Instituto de F\'{i}sica, Benem\'{e}rita Universidad Aut\'{o}noma
  de Puebla, Apartado Postal J-48, Puebla 72570, Mexico}
\affiliation{Dept. of Physics and Astronomy, Michigan State University, E. Lansing, Michigan 48824-1321, USA}

\maketitle

\onecolumngrid
\vspace*{0.4cm}

\begin{center}

{\large \bf Supplemental Material: \\ Emergence of correlations in the process of thermalization of interacting bosons }\\

\vspace{0.6cm}

Fausto Borgonovi$^{1,2}$, Felix M. Izrailev$^{3,4}$ \\

$^1${\it Dipartimento di Matematica e
  Fisica and Interdisciplinary Laboratories for Advanced Materials Physics,
  Universit\`a Cattolica, via Musei 41, 25121 Brescia, Italy}

$^2${\it Istituto Nazionale di Fisica Nucleare,  Sezione di Pavia,
  via Bassi 6, I-27100,  Pavia, Italy}
  
$^3${\it Instituto de F\'{i}sica, Benem\'{e}rita Universidad Aut\'{o}noma
  de Puebla, Apartado Postal J-48, Puebla 72570, Mexico}
  
$^4${\it Dept. of Physics and Astronomy, Michigan State University, E. Lansing, Michigan 48824-1321, USA}

\end{center}

%\vspace{0.6cm}

\section{Dynamics }

\noindent
Let us  consider initially an unperturbed  many-body
state of $H_0$, 
\begin{equation}
\ket{\psi(0)} = \ket{k_0} = \sum_\alpha C_{k_0}^\alpha \ket{\alpha} \ ,
\end{equation}
whose evolution under the Hamiltonian $H=H_0+V$ is given by
\begin{equation}
\langle k |\psi (t) \rangle = \langle k |e^{-iHt}| \psi (0) \rangle = \langle k |e^{-iHt}|k_0 \rangle   = \sum_\alpha C_{k_0}^\alpha C_{k}^\alpha e^{-iE^\alpha t},
\end{equation}
(note that all $C_k^\alpha$ are real numbers).
The probability to be in the unperturbed many-body state  $\ket{k}$ is 
\begin{equation}
P_k(t) = |\langle k |\psi (t) \rangle |^2 =  \sum_{\alpha,\beta}  C_{k_0}^\alpha C_{k}^\alpha 
C_{k_0}^\beta C_{k}^\beta 
e^{-i(E^\beta-E^\alpha )t} \ ,
\end{equation}
which can be written as a diagonal (time independent) plus a fluctuating (time-dependent) part,
\begin{equation}
\label{sdsf}
%\begin{array}{lll}
 P_k(t) =  \sum_{\alpha}  |C_{k_0}^\alpha|^2 |C_{k}^\alpha|^2
+  \sum_{\alpha\ne \beta}  C_{k_0}^\alpha C_{k}^\alpha 
C_{k_0}^\beta C_{k}^\beta 
e^{-i(E^\beta-E^\alpha )t} 
\equiv P_{k,k_0}^d + P_{k,k_0}^f (t).
%\end{array}
\end{equation}
Let us now define the long-time average of an observable $ A(t)$ as
\begin{equation}
\label{qta}
\overline{A} =   \lim_{T\to\infty} \ \frac{1}{T} \int_0^T \ dt \ A(t) .\ 
\end{equation}
It is  clear that for a non-degenerate spectrum  $\overline{P_{k, k_0}^f (t)}=0$ so that,
\begin{equation}
\overline{ P_k(t) } =  \sum_{\alpha}  |C_{k_0}^\alpha|^2 |C_{k}^\alpha|^2
  = P_{k, k_0}^d \ .
\end{equation}

%%%%%%%%%%%%%%%%%%%%%%%%%%%%%%%%%%%%%%%%%%%%%%%%%%%%%%%%%%%%%%%%%%%%%%%%%%%%%%%%
\begin{figure}[ht!]
\vspace{0.cm}
\includegraphics[scale=0.44]{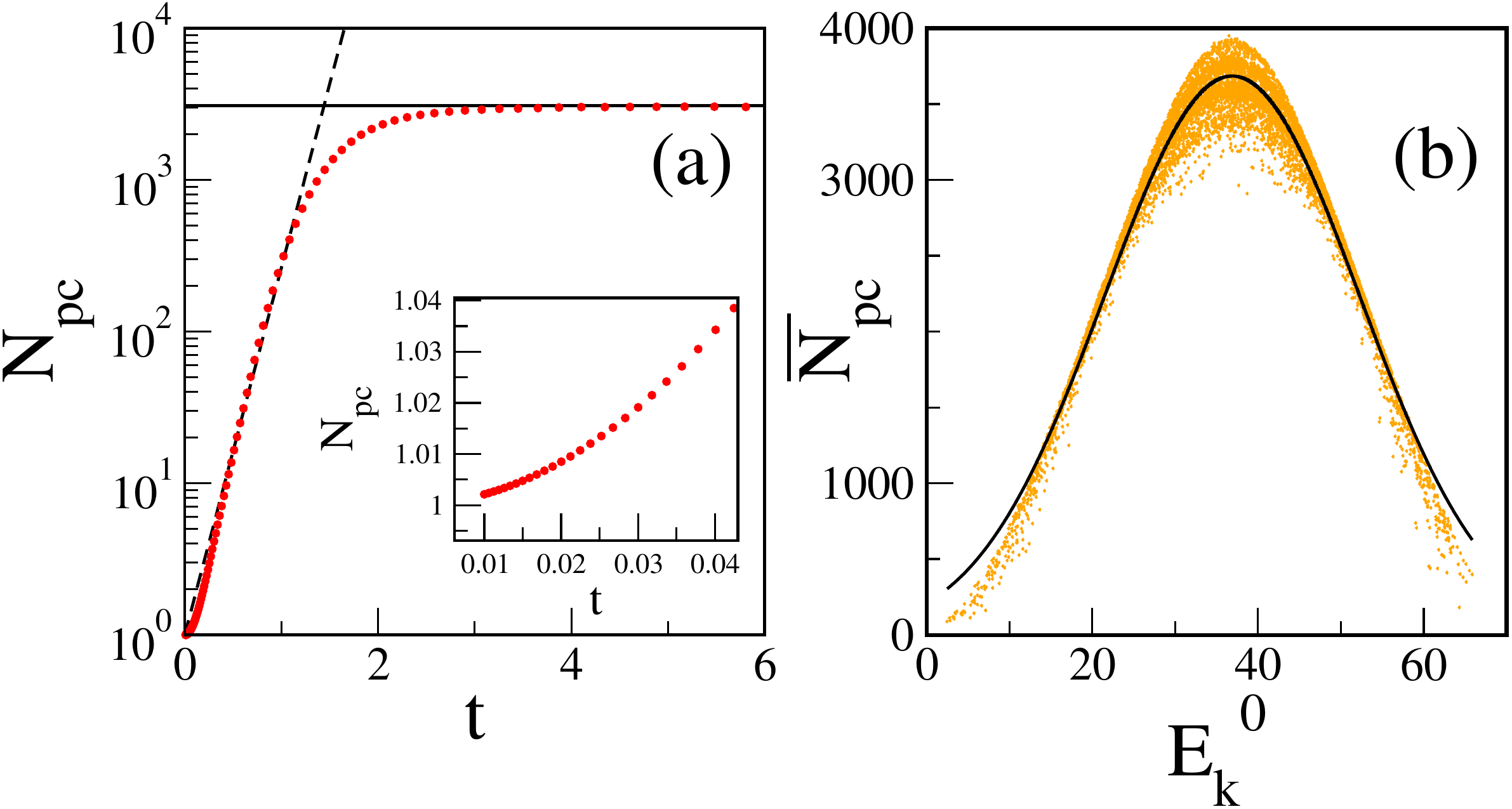}
\caption{(a) Number $N_{pc}(t)$ of principal components in time (red circles). Dashed line is the exponential growth with the rate $2\Gamma$ where $\Gamma \approx 2.8$ is the width of the LDOS found numerically from the decay of survival probability (for details see \cite{bis18}). Horizontal line is the estimate (\ref{t1-ipr}).
Inset : Initial  quadratic dependence $N_{pc} (t) \propto t^2$.
Initially all bosons are placed on the $5$-th single-particle level so that  $\ket{\psi_0} = \ket{k_0} = |0 0 0 0 6 0 0 0 0 0 0\rangle$.
(b) Orange dots represent the long-time average number of principal components as a function of the energy $E_k^0$ of the initial 
many-body  state. Black curve is a Gaussian fit. 
  Here is $N=6$, $M=11$, $V=0.4$.}  
\label{f-02}
\end{figure}
%%%%%%%%%%%%%%%%%%%%%%%%%%%%%%%%%%%%%%%%%%%%%%%%%%%%%%%%%%%%%%%%%%%%%%%%%%%%%%%%
\subsection{Number of Principal Components}

The long-time average for the number of principal
components can be computed as follows. Let us start from its definition,  
 \begin{equation}
\label{s-ipr-sm}
[N_{pc} (t)]^{-1} = \sum_k |\langle k |\psi (t) \rangle |^4  \equiv \sum_k [P_{k,k_0}^d + P_{k,k_0}^f (t)]^2.
\end{equation}
Taking the infinite-time average we have
 \begin{equation}
\label{t1-ipr}
[\overline{N}_{pc}]^{-1} = \sum_k (P_{k,k_0}^d)^2  + \overline{[P_{k,k_0}^f (t)]^2}.
\end{equation}
The second term in the r.h.s. of Eq.~(\ref{t1-ipr}) can be computed exactly,
 \begin{equation}
\label{t-sf2}
  \overline{[P_{k,k_0}^f (t)]^2} = (P_{k,k_0}^d)^2 -\sum_{\alpha}  \left| C_{k_0}^\alpha
 \right|^4 |C_{k}^\alpha|^4
\end{equation}
so that the long-time average for the number of principal components is given by,
\begin{equation}
\label{t-ipr-fin}
\overline{N}_{pc}  = \left[
2 \sum_k 
(P_{k,k_0}^d)^2  -  \sum_{\alpha}  | C_{k_0}^\alpha |^4 \sum_k |C_{k}^\alpha|^4
\right]^{-1}.
\end{equation}
This expression determines the  asymptotic value reached by $N_{pc}(t)$ after relaxation. It is shown in Fig.~\ref{f-02}(a) as a
horizontal line.
In the same figure we can identify three different regimes : a perturbative one for short time $t \ll 1/\Gamma$ where 
$N_{pc}(t)$ grows quadratically (see inset in Fig.~\ref{f-02} (a)); 
a second one characterized by the exponential growth, $N_{pc}(t) \simeq \exp(2\Gamma t)$ for $1/\Gamma \lesssim  t \lesssim N/\Gamma$, and a third one (saturation
after relaxation)
where $N_{pc}(t) \simeq  \overline{N}_{pc}$ for $t > N/\Gamma$ (for details see \cite{bis18}).

Another important information is how the stationary value $\overline{N}_{pc}$ depends on the initial state.
In Fig.~\ref{f-02}(b) we show $\overline{N}_{pc}$ as a function of the unperturbed energy $E_k^0$ of the initial many-body state $\ket{k_0}$.
As one can see it is quite well approximated (excluding the tails) by a Gaussian shape
(see black full curve).

\subsection{Single-particle Occupation Numbers}

Time dependent single-particle occupation numbers are defined as, 
\begin{equation}
\label{ond-1-sm}
n_{s} (t)  = \bra{\psi(t)} \hat{n}_s  \ket{\psi(t)} =  \sum_ k n_s^k  |\langle k |\psi (t) \rangle |^2 .
\end{equation}
Performing the infinite time average one obtains for the first two moments, 
\begin{equation}
\begin{array}{lll}
\overline{n_{s}} &=  \sum_k n_s^k  \overline{|\langle k |\psi (t) \rangle |^2 } = 
\sum_k n_s^k P_{k,k_0}^d \ \\
&\\
\overline{n_{s}^2} &=  \sum_k (n_s^k)^2  \overline{|\langle k |\psi (t) \rangle |^2 } = 
\sum_k (n_s^k)^2  P_{k,k_0}^d \ .
\label{noc-as2}
\end{array}
\end{equation}
and from that 
\begin{equation}
\delta n_{s}^2 (k_0) = 
\sum_k (n_s^k)^2  P_{k,k_0}^d  - \left(\sum_k n_s^k  P_{k,k_0}^d\right)^2 ,
\label{noc-asd}
\end{equation}
where the dependence on $k_0$ has been explicitly indicated in 
Eq.~(\ref{noc-asd}).

%%%%%%%%%%%%%%%%%%%%%%%%%%%%%%%%%%%%%%%%%%%%%%%%%%%%%%%%%%%%%%%%%%%%%%%%%%%%%%%% 
\begin{figure}[t]
\vspace{0.cm}
\includegraphics[scale=0.44]{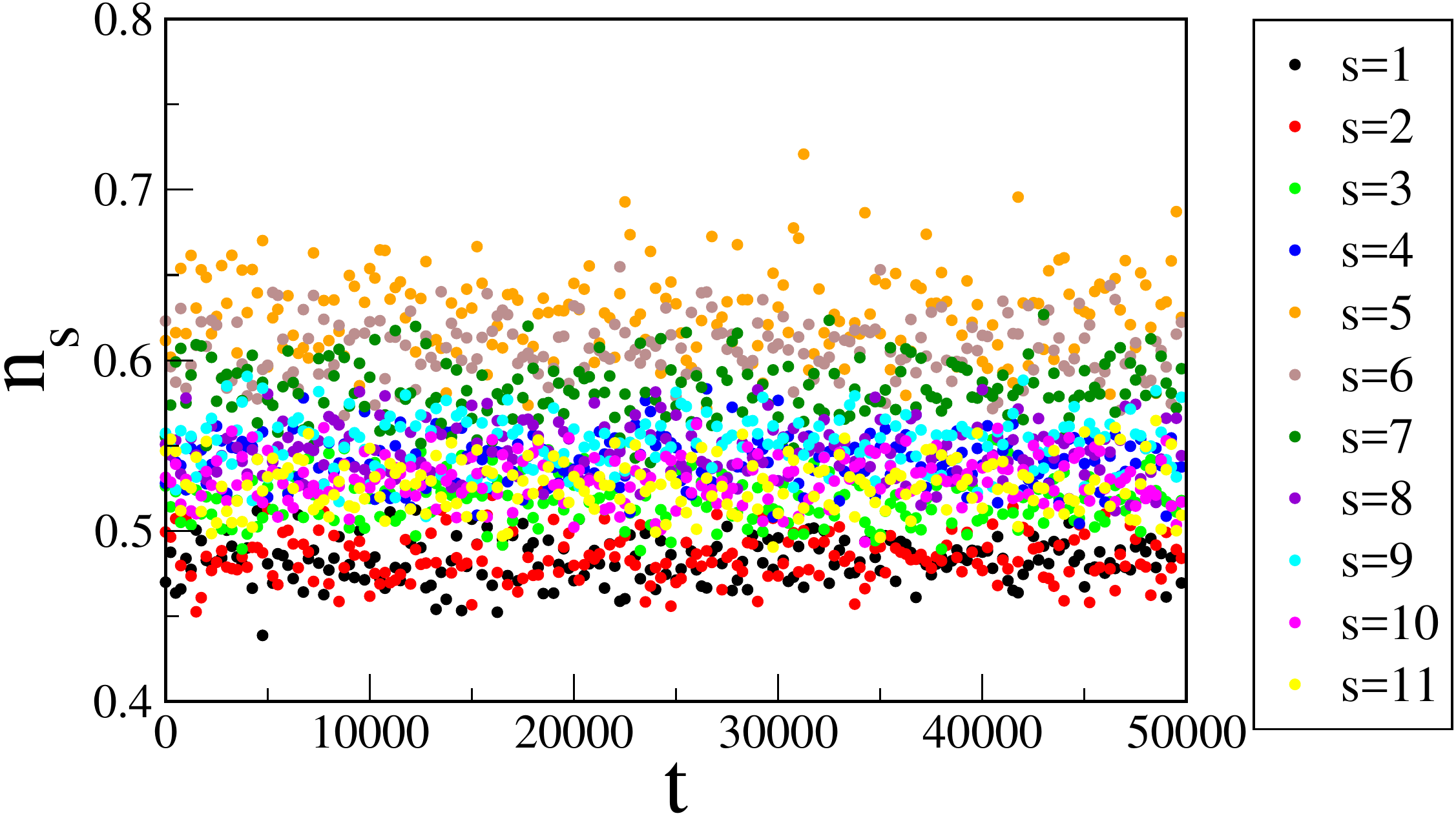}
\caption{Single-particle occupation numbers as a function of time after relaxation.
Different colors stand for different $s=1,..,M$.
Initial state is, in second quantized form,  $\ket{\psi_0} = |0 0 0 0 6 0 0 0 0 0 0\rangle$. Here is $N=6, M=11, V=0.4$.
}  
\label{f-055}
\end{figure}
%%%%%%%%%%%%%%%%%%%%

\subsection{Two-point Correlation Function}
First of all let us notice that the number operator $\hat{n}_s$ giving the number of particles in the 
single-particle energy level $\epsilon_s$ is diagonal in the unperturbed many-body basis, i.e.
\begin{equation}
\label{eq-d}
\bra{k} \hat{n}_s \ket{k'} = \delta_{k,k'} n_s^k.
\end{equation}
Concerning the global two-point correlation function one has, starting from the initial state $\ket{k_0}$,
\begin{equation}
\begin{array}{lll}
{\cal C}^{(2)}(t) &= \sum_{s=1}^{M-1}   \langle k_0 | [ \hat{n}_s(t) - \hat{n}_{s} ][ \hat{n}_{s+1}(t) - \hat{n}_{s+1}]|  k_0 \rangle  \\
&\\
& = \sum_{s=1}^{M-1} \langle k_0 |  \hat{n}_s(t)\hat{n}_{s+1}(t)|  k_0 \rangle -n_s^{k_0} \langle k_0|  n_{s+1}(t)|k_0 \rangle - 
n_{s+1}^{k_0} \langle k_0|  n_{s}(t)| k_0 \rangle  + n_s^{k_0} n_{s+1}^{k_0} \\
&\\
&= \sum_{s=1}^{M-1}     \sum_k  | \langle k | \psi(t) \rangle |^2  [n_s^k n_{r}^k +  n_s^{k_0} n_{r}^{k_0} -n_s^{k_0} n_{r}^{k}-n_s^{k} n_{r}^{k_0} ] \equiv \sum_{s=1}^{M-1}    \sum_k  | \langle k | \psi(t) \rangle |^2  W_{k,k_0}^{sr},
\end{array}
 \label{corr11a}
\end{equation}
where $\hat{n}_s(t) = e^{iHt} \hat{n}_s e^{-iHt} $. In Eq.~(\ref{corr11a}) 
 we have defined 
\begin{equation}
  W_{k,k_0}^{sr} =  [n_s^k n_{r}^k +  n_s^{k_0} n_{r}^{k_0} -n_s^{k_0} n_{r}^{k}-n_s^{k} n_{r}^{k_0} ].  
 \label{corr11b}
\end{equation}
The long-time average is thus given by, 
 \begin{equation}
\overline{{\cal C}^{(2) }}   =   \sum_{s=1}^{M-1}    \sum_k  P_{k,k_0}^d  W_{k,k_0}^{sr}.
\label{cor-ave-sm}
\end{equation}

\subsection{Four-point Correlation Function}
Let us obtain the long-time estimate for the four-point correlation function (OTOC): 
\begin{equation}
{\cal O}_{s,s+1} (t) = \langle k_0 |[ \hat{n}_s(t), \hat{n}_{s+1}(0)] |^2 | k_0 \rangle .
\label{otoc1-sm}
\end{equation}
From the definition it is clear that ${\cal O}_{s,s+1} (0)=0$.
In order to compute explicitly Eq.~(\ref{otoc1-sm}) let us insert a completeness so that,
\begin{equation}
{\cal O}_{s,s+1} (t) = \sum_k  | \langle k_0 | \hat{n}_s(t) | k \rangle |^2 \left( n_{s+1}^k - n_{s+1}^{k_0}\right) ^2  .
\label{otoc1a}
\end{equation}
Setting
 \begin{equation} 
 \bra{k_0} \hat{n}_s(t) \ket{k}  = \sum_q {\cal F}_{k,q} (t)  {\cal F}_{k_0,q}^* (t) n_s^q \ ,
\label{otoc2} 
\end{equation}
where we have defined
 \begin{equation} 
  {\cal F}_{k,q} (t)   = \bra{q} e^{-iHt} \ket{k} = \sum_{\alpha} C_q^\alpha C_k^\alpha e^{-iE^\alpha t},
\label{def-f} 
\end{equation}
the long-time average can be written as
 \begin{equation}
\overline{{\cal O}_{s,s+1}  } =    \sum_k  
\left(n_{s+1}^k -n_{s+1}^{k_0}\right)^2
\left\{  \left[\sum_{\alpha} C_k^\alpha  C_{k_0}^\alpha {\cal N}_s^{\alpha,\alpha}\right]^2+
\sum_{\alpha \ne \beta} |C_k^\alpha|^2  |C_{k_0}^\beta|^2 \left( {\cal N}_s^{\alpha,\beta} \right)^2,
\right\}
\label{otoc-ave-sm}
\end{equation}
where we have defined, for each $s$,  the matrix
 \begin{equation}
 {\cal N}_s^{\alpha,\beta} = 
\sum_{k} C_k^\alpha  C_{k}^\beta n_s^k.
\label{otoc-s}
\end{equation}

%%%%%%%%%%%%%%%%%%%%%%%%%%%%%%%%%%%%%%%%%%%%%%%%%%%%%%%%%%%%%%%%%%%%%%%%%%%%%%%%
\begin{figure}[t]
\vspace{0.cm}
\includegraphics[scale=0.44]{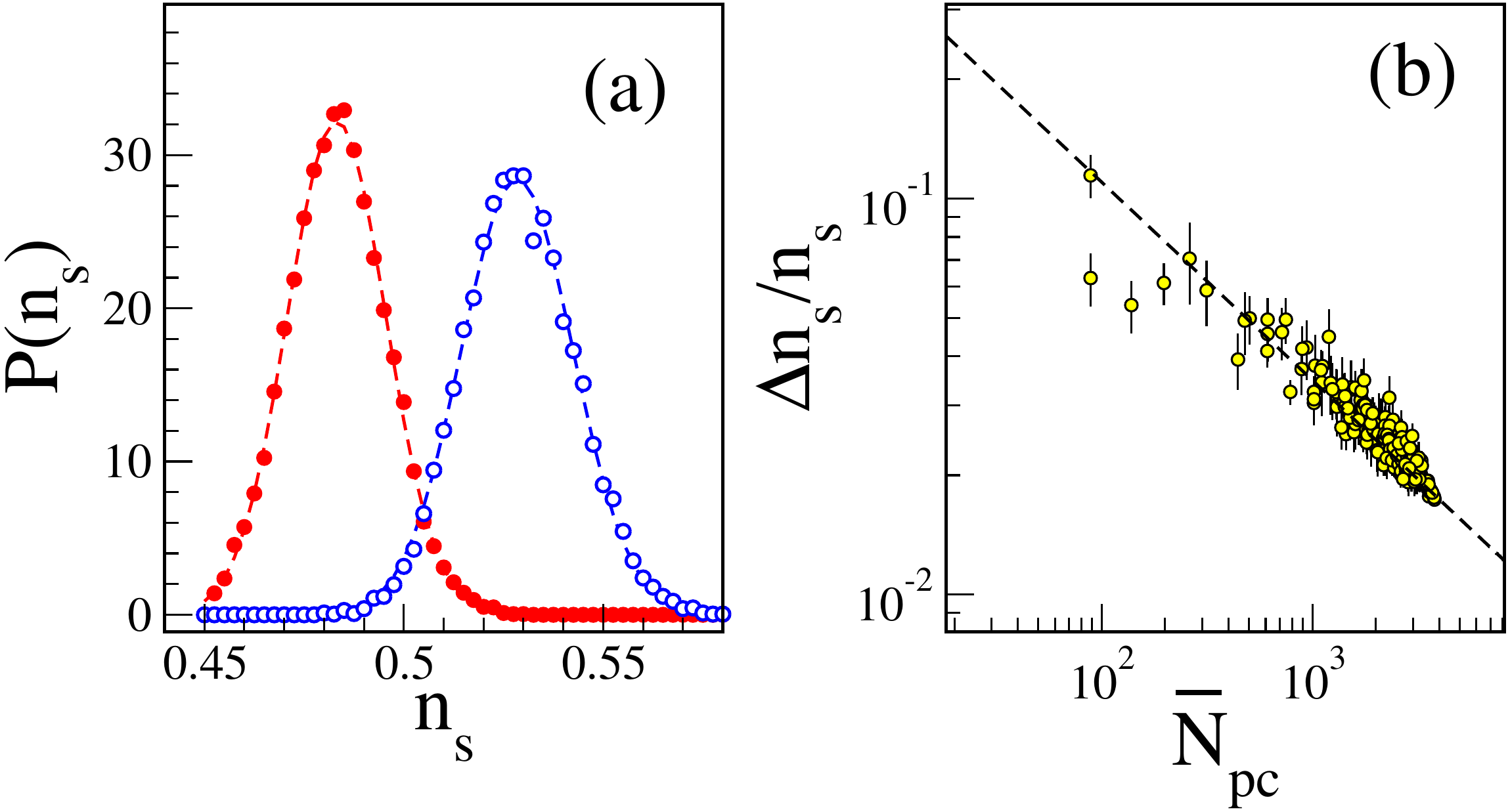}
\caption{(a) Probability distribution $P(n_s)$ for two different $s$ values : $s=1$ (full red symbols) and $s=11$ (open blue symbols).
Data are obtained from Fig.~\ref{f-055}. Dashed lines represent fits with Gaussian distributions.  (b)
 : Relative time-fluctuations $\Delta n_s/n_s$   as a function of the correspondent number of principal 
components $\overline{N}_{pc}$ obtained from the stationary distribution. Dashed line is $1/\sqrt{\overline{N}_{pc}}$.
Data are $N=6, M=11, V=0.4$. 
}
\label{f-06}
\end{figure}
%%%%%%%%%%%%%%%%%%%%%%%%%%%%%%%%%%%%%%%%%%%%%%%%%%%%%%%%%%%%%%%%%%%%%%%%%%%%%%%%

\section{Classical and quantum  Fluctuations}
In this section we study the statistical properties of the stationary distribution of single-particle occupation numbers.
  In particular we analyze both ``classical'' and
``quantum'' fluctuations.
Concerning the former they can be obtained from the study of the time fluctuations of $n_s(t)$ around its 
infinite time average.
Statistical relaxation should be characterized by small
  fluctuations of $n_s$    compared with the mean values $\langle n_s \rangle$, and  of Gaussian type.  

In Fig.~\ref{f-055} the long-time dynamics of the average occupation numbers 
$$n_s(t)= \langle k_0|\hat{n}_s(t) |k_0\rangle$$ are shown for different $s$ values.
Let us first concentrate on  the statistical
properties of this ``classical signal'', $n_s(t)$. The distributions $P(n_s)$, taken from the values in Fig.~\ref{f-055} are shown
 in Fig.~\ref{f-06}(a) (for two values of $s$:    $s=1$  and $s=M$).
As one can see there is a very good agreement with a Gaussian fit.
The width of these distributions (as given by the second moment of the fitted Gaussians $\Delta n_s^2$)  weakly depends on
 the particular chosen $s$ value
(see Fig.~\ref{f-06}(a)) while the dependence on the initial state is stronger.
To this end we compute the relative fluctuations $\Delta n_s/n_s $ choosing as initial states different unperturbed 
many-body basis states from the whole energy spectrum. 
In agreement what the results found for Fermi and Bose  particles~\cite{fi97,sm-BMI17}, we consider in Fig.~\ref{f-06}(b) the relative fluctuations 
$\Delta n_s/n_s $
as a function of the number of principal components of the stationary wave-packet  (after relaxation) for the correspondent initial states
(essentially what is shown in Fig.~\ref{f-02}(b).)
As one can see there is a very good agreement with the dependence $1/\sqrt{\overline{N}_{pc}}$
which is a strong result in view of the requirement of statistical mechanics. Let us  stress that the decrease of relative fluctuations occurs not with respect to the number  $N$ of particles, but   with the number of principal components contained in the stationary distribution $\overline{N}_{pc}$.
 
\begin{figure}[t]
\vspace{0.5cm}
\includegraphics[scale=0.44]{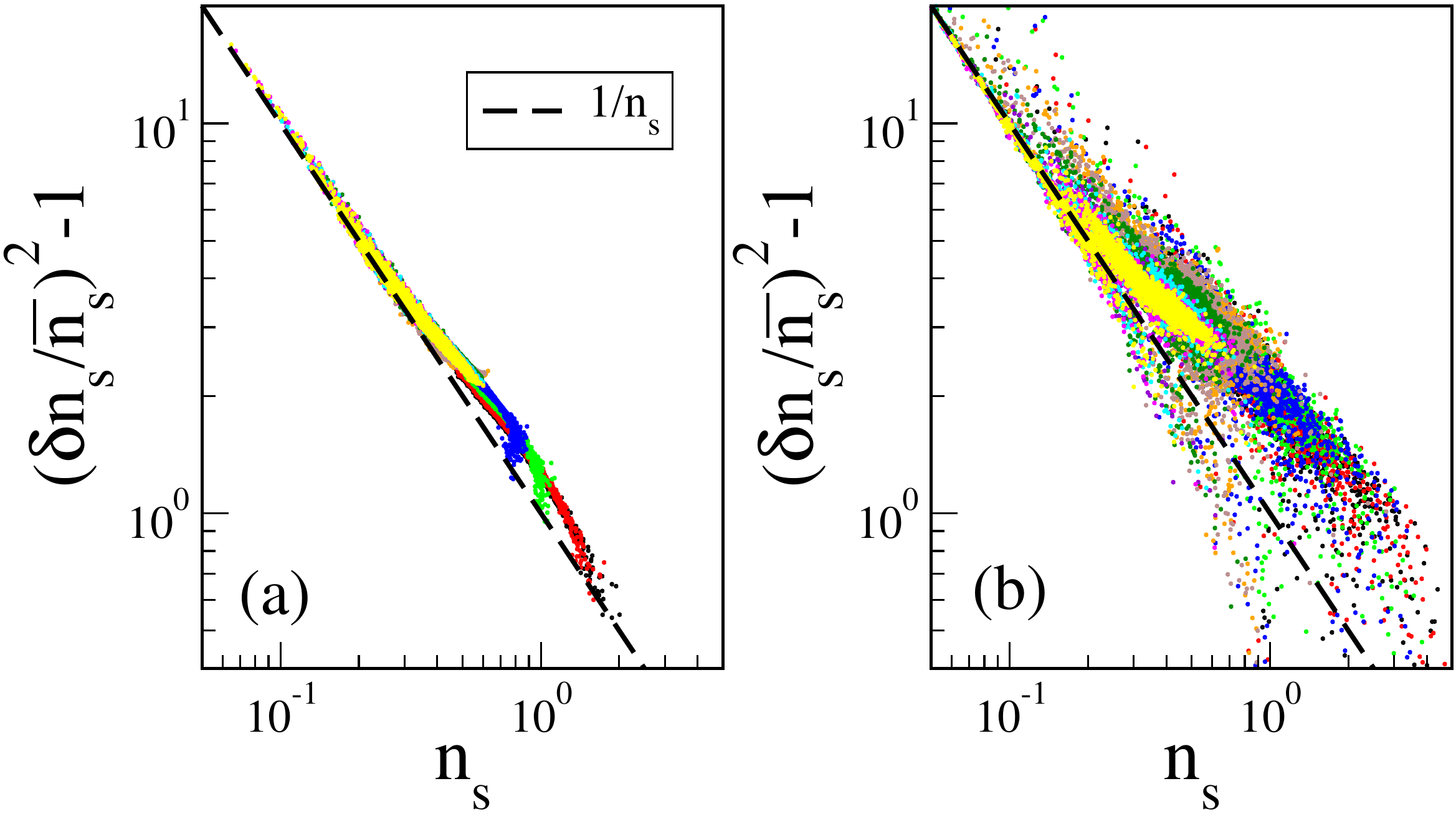}
\caption{Relative quantum fluctuations $(\delta n_s/n_s)^2-1$. Initial states  $\ket{k_0}$ are   basis states chosen in the whole energy spectrum. On $x$-axis the averaged values of $n_s$ are plotted. Dashed line is the theoretical prediction $1/n_s$. Different colors refer to different $s$ values. (a) $V=0.4$ case of strong quantum chaos, (b) $V=0.04$ case of  non chaotic eigenstates for which 
Eq.~(\ref{eq-qfl1}) is not valid.
}
\label{f-08}
\end{figure}

Concerning  quantum fluctuations, they are defined by 
\begin{equation}
\label{eq-qfl}
\delta n_s^2 (k_0) =    \overline{n_{s}^2} - (\overline{ n_{s}})^2        
\end{equation}
for different initial states $\ket{k_0}$, 
and from them, the relative fluctuations $\delta n_s/\overline{ n_{s}}$. In the canonical ensemble, for {\it non-interacting} bosons   the following relation holds \cite{pathria},
\begin{equation}
\label{eq-qfl1}
\frac{\delta n_s^2}{n_s^2} = 1+ \frac{1}{n_s}
\end{equation}
We have numerically checked this relation, see data in Fig.\ref{f-08}(a) from which one can see a good correspondence to the above relation in the case when the eigenstates are strongly chaotic. 
In   Fig.\ref{f-08}(b), the same quantity has been plotted   for a non-chaotic case.
As one can see   quantum fluctuations deviate strongly from the prediction given in Eq.~(\ref{eq-qfl1}).
This result shows once more that even for a finite number of particles, provided a  strong enough inter-particle interaction,  conventional statistical mechanics works extremely well.

\end{document}